\documentclass[prx,superscriptaddress,twocolumn]{revtex4}

\usepackage[SquareTraceBrackets]{quantum}
\usepackage{graphicx,bm,natbib,upgreek,amsbsy,amsmath,mathrsfs, dsfont, upgreek}
\usepackage[dvipsnames]{color}
\definecolor{myblue}{named}{MidnightBlue}
\definecolor{mygreen}{RGB}{0,120,0}
\usepackage[colorlinks=true,citecolor=myblue,linkcolor=myblue,urlcolor=myblue]{hyperref}
\usepackage[T1]{fontenc}
\usepackage{newtxtext,newtxmath}
\usepackage[scaled]{helvet}
\usepackage{dcolumn}		

\DeclareGraphicsExtensions{.pdf, .jpg, .eps, .svg}

\begin{document}

\title{Quantum metrology of spatial deformation using arrays of classical and quantum light emitters}
\author{Jasminder S. Sidhu}
\email{jsidhu1@sheffield.ac.uk}
\author{Pieter Kok}%
\email{p.kok@sheffield.ac.uk}
\affiliation{%
Department of Physics and Astronomy, The University of Sheffield, Sheffield, S3 7RH, UK.}
\date{\today}

\begin{abstract}

We introduce spatial deformations to an array of light sources and study how the estimation precision of the interspacing distance, $d$, changes with the sources of light used. The quantum Fisher information (\textsc{qfi}) is used as the figure of merit in this work to quantify the amount of information we have on the estimation parameter. We derive the generator of translations, $\mathcal{\hat{G}}$, in $d$ due to an arbitrary homogeneous deformation applied to the array. We show how the variance of the generator can be used to easily consider how different deformations and light sources can effect the estimation precision. The single parameter estimation problem is applied to the array and we report on the optimal state that maximises the \textsc{qfi} for $d$. Contrary to what may have been expected, the higher average mode occupancies of the classical states performs better in estimating $d$ when compared with single photon emitters (\textsc{spe}s). The optimal entangled state is constructed from the eigenvectors of the generator and found to outperform all these states. We also find the existence of multiple optimal estimators for the measurement of $d$. 
Our results find applications in evaluating stresses and strains, fracture prevention in materials expressing great sensitivities to deformations, and selecting frequency distinguished quantum sources from an array of reference sources. 
\end{abstract}

\maketitle

\section{Introduction}
\label{sec:label}

Quantum mechanics has established physical limitations to precision bounds in a myriad of applications in parameter estimation~\cite{Bollinger1996_PRA, Wolfgramm2012_NP, Berrada2013_NC, Napolitano2011_N}. Approaching these limitations through high precision measurements is one of the principal objective in quantum metrology. These efforts have seen the development of fundamental theories across science. An immediate exemplification of this can be recognised in the measurement of gravitational waves. Exotic states of light, such as squeezed light, is now routinely used to enhance the sensitivities of large interferometers, such as \textsc{ligo} and \textsc{virgo} for the measurement of gravitational waves~\cite{Ligo2013_NP, Abbott2004_PRD, Caron1995_NI}. Estimations of gravitational wave amplitudes have also been made by considering phonons in Bose-Einstein condensates~\cite{Carlos2014_NJP}. Besides this, quantum enhanced measurements have fruitfully demonstrated performance improvements in atomic clocks, remote sensing, navigation, and thermometry~\cite{Toth2014_JPA, Degen2016_arxiv, Pearce2016_arxiv}.

Quantum metrology is rooted in the theory of quantum parameter estimation, pioneered by Holevo~\cite{Holevo2011} and Helstrom~\cite{Helstrom1976}. The quantum Cram\'{e}r-Rao bound (\textsc{qcrb}) has become a standard tool in providing a lower bound on the variance of an unbiased estimator that maps measured data from quantum measurements to parameter estimations. It provides a fundamental bound to the achievable precision of any estimating strategy and is intrinsically dependent on only the uncertainty inherent in the quantum state. Clearly any performance improvements in parameter estimations then manifests itself in our ability to manipulate the quantum nature of light. More meaningfully, the precision of parameter estimates is bounded by the physical resources, a matter addressed by the query complexity of the quantum network describing the estimation procedure~\cite{Zwierz2012_PRA}. It is well known that quantum resources provide improvements to the estimation sensitivities of physical parameters and this has been demonstrated for phase estimations in interferometers~\cite{Giovannetti_Metrology_paper_2006, Kok2010, Dobrzanski2009_PRA, Pezze2014_AI}, and such resources are exploited in quantum imaging to drive resolution capabilities past the Abbe-Rayleigh criterion~\cite{Abbe1873, Rayleigh1879, Oppel2012_PRL, Pearce2015_PRA, Thiel2007_PRL, Abouraddy2001_PRL, Fei1997_PRL, Kolobov2007, Boto2000_PRL, Lloyd2008_S}.

A large proportion of metrological protocols can be reduced to that of phase estimation~\cite{Giovannetti2004_S}. Extensive studies have thus been made in estimating multiplicative factors of Hamiltonians, given their enhancement to phase and frequency estimations. A generalisation of this to arbitrary Hamiltonian parameters was addressed in~\cite{Pang2014_PRA}. Further advances include parameter estimation of dissipative dynamics~\cite{Alipour2014_PRL, Monras2011_PRA, Braun2011_NC}, nuclear properties of spin $1/2$ chains~\cite{Marzolino2014_PRA} and magnetic field measurements~\cite{Baumgratz2016_PRL}. We focus instead on directly estimating distances between neighbouring light sources along an array. We evaluate how changing the nature of the light sources attached and array deformations can impact the estimation precision. This is essential in detecting stresses and strains exerted on materials with great sensitivities to deformations, and allows for corrective measures to negate the effects to prevent fractures. It may also precisely determine the coordinates of the emitters and select particular sources distinguished by its frequency from an array of reference or differing sources~\cite{Tsang2014_O}.

We start by reviewing quantum estimation theory and introduce the quantum Fisher information (\textsc{qfi}) in Sec. \ref{sec:qfi}. Specifically, we outline the \textsc{qfi} for unitary transformations of pure and mixed states and how it relates to the generator of translations $\mathcal{\hat{G}}$ in the estimating parameter. This formulation is particularly convenient since the operator is independent of the choice of initial states such that the \textsc{qfi} is determined only by $\mathcal{\hat{G}}$ and the initial state. In Sec.~\ref{sec:generator} we derive the form of the generator of translations in the source separation distance $d$ of a stationary array of arbitrary sources due to some general applied homogeneous deformation $\bm{\Xi}$. We apply the generator in Sec.~\ref{sec:diff_sources} to capture the dynamics of the state parameterisation after a stationary 1-dimensional array of classical and quantum light sources undergoes a stretching deformation $\bm{\Xi} \rightarrow \xi_s$. The parameterisation arises from the parings of different sources along the array. We calculate the \textsc{qfi} to compare the performance of arrays of single photon emitters (\textsc{spe}s), coherent, thermal, and entangled sources of light on the estimation of $d$. In contrast to what may have been expected from earlier work~\cite{Oppel2012_PRL, Thiel2007_PRL}, we find that the higher mode occupancies of classical coherent and thermal states affords better estimation precisions when compared with the \textsc{spe}s. This would be favourable since generating classical states may be less resource-expensive to create. However a quantum enhancement is observed when entanglement is employed. In agreement with separate work, the optimal state is that which entangles the eigenstates corresponding to the maximum and minimum difference eigenvalues of the generator. We demonstrate that not all entangled states can reproduce similar precision enhancements. This insight is reminiscent of previous studies where entanglement was concluded a necessary but insufficient resource for quantum metrology~\cite{Tsang2008_PRL, Gottesman1997_Thesis, Tilma2010_PRA, Braun2017_arxiv}. The all these studies, rarely are the optimal measurement strategies considered. To address this, we discuss the optimal estimator for \textsc{spe}s in Sec.~\ref{sec:optimal_states} and report our conclusions in Sec.~\ref{sec:conclusions}. Appendix materials have been provided.


\section{The quantum Fisher information}
\label{sec:qfi}

The \textsc{qfi} quantifies the amount of information about a parameter in a state. It is a property of the state and does not depend on the measurement strategy. For a vector of parameters, $\bm{\vartheta} = (\vartheta_1, \vartheta_2, \ldots, \vartheta_M)^\top$, the \textsc{qfi} for the state
\begin{align}
\smash{\rho(\bm{\vartheta}) = \sum_j p_j(\bm{\vartheta}) \vert\varrho_j(\bm{\vartheta})\rangle \langle\varrho_j(\bm{\vartheta})\vert}
\label{eqn:general_state}
\end{align}
provides a lower bound on the average mean-square error (\textsc{mse}). This is the quantum Cram\'{e}r-Rao bound (\textsc{qcrb})~\cite{Holevo2011, Helstrom1976}
\begin{align}
\left[\text{Cov}\left(\bm{\vartheta}\right)\right]_{jk} \geq \frac{1}{4\nu}\left[(\bm{\mathcal{I}}^Q)^{-1}\right]_{jk},
\label{eq:qfi_sld_1}
\end{align}
where $\text{Cov}\left(\bm{\vartheta}\right)$ is the covariance matrix of $\bm{\vartheta}$ and $\nu$ the number of independently repeated measurements. Defining $\smash{\bm{\mathcal{L}} = (\mathcal{L}_1, \mathcal{L}_2, \ldots, \mathcal{L}_M)^\top}$, the matrix elements of the \textsc{qfi} in Eq. \eqref{eq:qfi_sld_1} may be written in terms of the symmetric logarithmic derivatives \textsc{(sld)}, $\bm{\mathcal{L}}_j$, as
\begin{align}
\left[\bm{\mathcal{I}}^Q\right]_{jk} = \frac{1}{2}\Tr{\rho(\bm{\vartheta}) \{\bm{\mathcal{L}}_j, \bm{\mathcal{L}}_k\}}.
\label{eq:qfi_sld_2}
\end{align}
The \textsc{sld}, defined implicitly by 
\begin{align}
2\partial_j\rho(\bm{\vartheta}) = \left\{\rho(\bm{\vartheta}), \bm{\mathcal{L}}_j\right\},
\label{eqn:sld_implicit_definition}
\end{align} 
where $\smash{\partial_j = \partial/\partial\vartheta_j}$ is the derivative with respect to the parameter $\vartheta_j$. A property of the \textsc{sld} is given by the trace of Eq.~\eqref{eqn:sld_implicit_definition}, whereupon we observe that the expectation value $\braket{\bm{\mathcal{L}}_j} = 0$ for arbitrary $\rho(\bm{\vartheta})$. In information geometry, the Fisher information metric is defined on a statistical manifold whose points, $P(\vartheta)$ are probability measures. Viewed in this way and defining the statistical distance $\delta s(\vartheta) = P(\vartheta + \delta \vartheta) - P(\vartheta)$ resulting from some unitary process on the state, then the Fisher information may be defined as the square of the rate of change of the statistical distance. It is meaningful then to understand how the \textsc{qfi} may be determined from the generator of translations in $\bm{\vartheta}$ due to some unitary process. The operator representation has a further advantage that it does not rely on a particular basis.

Consider some unitary process where the state parameterisation is introduced through the unitary $\hat{U}(\bm{\vartheta})$, then $\smash{\rho(\bm{\vartheta}) = \hat{U}(\bm{\vartheta}) \rho(0) \hat{U}^\dagger(\bm{\vartheta})}$ with $\hat{U}(\bm{\vartheta})^\dagger$ the transposed complex conjugate of $\hat{U}(\bm{\vartheta})$. For some generator of dynamics in the vector of parameters $\bm{\vartheta}$, $\hat{\bm{\mathcal{F}}}$, the unitary may be written as
\begin{align}
\hat{U}(\bm{\vartheta}) = \exp\left[- i \hat{\bm{\mathcal{F}}} \cdot \bm{\vartheta}\right].
\label{eqn:parameter_unitary}
\end{align}
The operator $\hat{\bm{\mathcal{F}}}$ is a local generator which characterises the sensitivity of the system state $\rho(\bm{\vartheta})$ on changes in $\bm{\vartheta}$ after unitary evolutions. If the unitary for the physical process governing the parameterisation is known, the generator of changes in $\vartheta_j$ is defined~\cite{Pang2014_PRA}
\begin{align}
\hat{\mathcal{F}}_j = i \hat{U}^\dagger(\bm{\vartheta}) \partial_j \hat{U}(\bm{\vartheta}),
\label{eqn:generator_form}
\end{align}
which is Hermitian since $\smash{\hat{U}^\dagger(\partial_d\hat{U}) = - (\partial_d\hat{U}^\dagger)\hat{U}}$ and can be easily demonstrated by Taylor expanding of $\rho(\bm{\vartheta})$. The matrix elements of the \textsc{qfi} as in Eq.~(\ref{eq:qfi_sld_2}) may be re-written in terms of the generator as~\cite{Liu2014_CTP}
\begin{align}
\begin{split}
\left[\bm{\mathcal{I}}^Q\right]_{mn} \hspace{-.2em}&= \hspace{-.2em}\sum_{j = 1}^D 4p_j \left[\text{Cov}\left(\hat{\mathcal{F}}_m, \hat{\mathcal{F}}_n\right)\right]_{j} \\
&- \hspace{-.2em}\sum_{j \neq k}^D \hspace{-.1em} \frac{8p_jp_k}{(p_j + p_k)}\Braket{\varrho_j\vert\hat{\mathcal{F}}_m\vert\varrho_k} \Braket{\varrho_k \vert\hat{\mathcal{F}}_n\vert\varrho_j}, 
\label{eqn:qfi_generator_form}
\end{split}
\end{align}
where $D = \text{dim}[\text{supp}(\rho(\bm{\vartheta}))]$ is the dimension of the support set of $\rho(\bm{\vartheta})$ and $\{m,n \in \mathbbm{Z} \vert 1 \leq m, n \leq D\}$ define the elements of the \textsc{qfi} matrix elements. The covariance matrix of the generators on the $j^\text{th}$-eigenstate of the initial state in Eq.~\eqref{eqn:qfi_generator_form} is defined as
\begin{align}
\begin{split}
\left[\text{Cov}\left(\hat{\mathcal{F}}_m, \hat{\mathcal{F}}_n\right)\right]_{j} & = \frac{1}{2} \braket{\varrho_j\vert\{\hat{\mathcal{F}}_m, \hat{\mathcal{F}}_n\}\vert\varrho_j} \\
& - \braket{\varrho_j\vert \hat{\mathcal{F}}_m\vert\varrho_j}\braket{\varrho_j\vert \hat{\mathcal{F}}_n\vert\varrho_j},
\label{eqn:covariance_matrix}
\end{split}
\end{align}
where $\{\cdot, \cdot\}$ defines the anticommutator. Eq.~\eqref{eqn:qfi_generator_form} is often written in the following compact form~\cite{Braunstein1994_PRL, Liu2015_SR}
\begin{align}
\left[\bm{\mathcal{I}}^Q\right]_{mn} \leq 4 \left[\text{Cov}\left(\hat{\mathcal{F}}_m, \hat{\mathcal{F}}_n\right)\right]_{\text{input}},
\label{eqn:qfi_generator_pure}
\end{align}
where the subscript `input' describes the initial input state. The equality is strictly limited to pure states and the bound for all other quantum states. We summarise some of the convenient subtleties that arise from the unitary transformation formulation of the \textsc{qfi}. First, the generator captures the dynamics of the parameterisation process of the state and is basis-independent. Second, the \textsc{qfi} depends only on the generator and the initial states. The form in Eq.~\eqref{eqn:qfi_generator_pure} provides an easily computable upper bound on the \textsc{qfi} for different quantum states. A zero generator variance results when the state is invariant under unitary dynamics of the type described by Eq.~\eqref{eqn:parameter_unitary}. Third, entanglement between specific eigenstates of the generator can be used to construct an optimal state which maximises the \textsc{qfi}~\cite{Giovannetti2011_NP}.
\begin{figure}[t!]
\centering
\includegraphics[width =0.95\columnwidth]{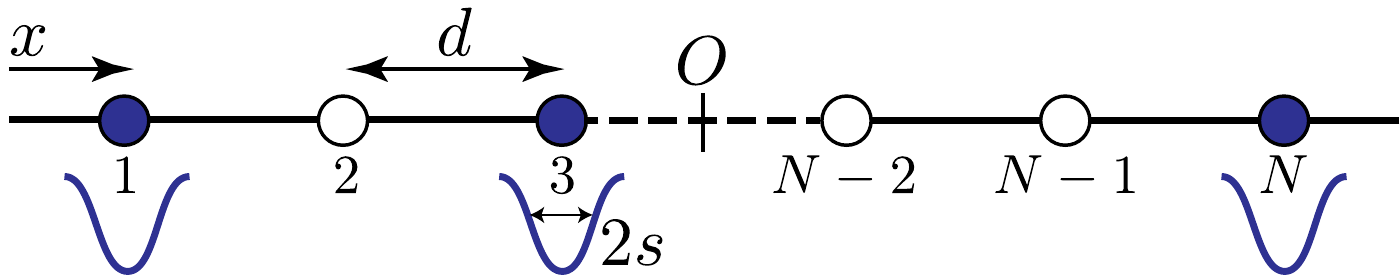}
\caption{(Colour online). Array of identical, stationary and equidistant emitters each with an intrinsic spatial Gaussian uncertainty $s$ (blue envelopes). The continuous variable $x$ runs along the source plane, $d$ is the source separation and $O$ defines the midpoint of the array of $N$ photons. For a source operation efficiency, $\eta$, filled sources are understood to emit a photon whereas the unfilled sources do not. In this work, all sources will be assumed to operate with efficiency $\eta = 1$.}
\label{pic:stationary_spe_chain}
\end{figure}%
Finally, since the \textsc{qfi} can be determined either from the generator of dynamics in the parameter to be estimated or directly from the \textsc{sld} through~\eqref{eq:qfi_sld_2}, both formulations are equivalent. To illustrate this, we relate the \textsc{sld} to the generator. The \textsc{sld} as defined in Eq.~\eqref{eqn:sld_implicit_definition} describes dynamics of the quantum state $\rho(d)$. Since unitary dynamics are generally given by the von-Neumann equation
\begin{align}
\partial_j\rho(\bm{\vartheta}) = i \left[\rho(\bm{\vartheta}),\hat{\mathcal{F}}_j\right],
\label{eqn:von_neumann_eqn}
\end{align}
then comparison with Eq.~\eqref{eqn:sld_implicit_definition} gives~\cite{Toth2014_JPA} 
\begin{align}
\bm{\mathcal{L}}_j = 2i \sum_{k, l = 1}^D \frac{[\hat{\mathcal{F}}_j]_{k, l}(p_k - p_l)}{p_k + p_l} \ket{\psi_k}\bra{\psi_l}
\label{eqn:sld_equation}
\end{align}
where $[\hat{\mathcal{F}}_j]_{k, l}$ defines the matrix elements of the generator.

Determining the \textsc{sld} addresses another important question in quantum metrology. The \textsc{qcrb} in Eq.~\eqref{eq:qfi_sld_1} defines the smallest achievable precision bound for parameter estimations, but does not define the optimal measurement which saturates it. This precision bound may be achieved by performing measurements in the eigenbasis of the \textsc{sld} operator, if its eigenvectors are locally independent of the parameter~\cite{Braunstein1994_PRL, Braunstein1996_AoP}. The optimal quantum estimator which saturates the \textsc{qcrb} for $\vartheta_j$ is given by
\begin{align}
\check O(\vartheta_j) = \vartheta_j \mathds{1} + \left[(\bm{\mathcal{I}}^Q)^{-1} \cdot \bm{\mathcal{L}}\right]_j,
\label{eqn:optimal_measure_operator}
\end{align}
which is a projective measurement onto the eigenstates of the \textsc{sld}~\cite{Paris2009_IJQI}. The first term represents the average estimate and the second the smallest covariance of the optimal measurement. Determining the measurement $\check O(\vartheta_j)$ is generally a difficult task since it depends on the parameter to be estimated, $\vartheta_j$. In such cases, better precision bounds may be achieved through adaptive measurements~\cite{Berry2002_PRA, Berry2000_PRL}. For single parameters the \textsc{qcrb} provides an ultimate bound for unbiased estimators, $\smash{\braket{\check{O}(\vartheta_j)} = \vartheta_j}$~\footnote{We use the caron to distinguish estimators from quantum mechanical operators.} and can be asymptotically saturated through maximum likelihood estimation~\cite{Braunstein1992_PRL}. For multiple parameters, the \textsc{qcrb} is generally not attainable for simultaneous measurements of each parameter if the \textsc{sld}s associated with the parameters do not commute. This generally makes the multivariate \textsc{qcrb} non-saturable and introduces a further theoretical complication above that of optimal measurements depending on the true values of the parameters. However, even for incompatible \textsc{sld} operators, the multiparameter \textsc{sld} \textsc{qcrb} remains asymptotically attainable if and only if~\cite{Gao2014_EPJD}
\begin{align}
\text{Tr}\left(\rho(\bm{\vartheta})[\bm{\mathcal{L}}_j, \bm{\mathcal{L}}_k]\right) = 0.
\label{eqn:asymp_attainability_criteria}
\end{align}
Differing methods to provide better precision bounds may involve collective measurements over many independent copies of the system, which is experimentally challenging.


\section{Generator of translations}
\label{sec:generator}

\begin{figure}[t!]
\begin{center}
\includegraphics[width =\columnwidth, height = 0.618\columnwidth]{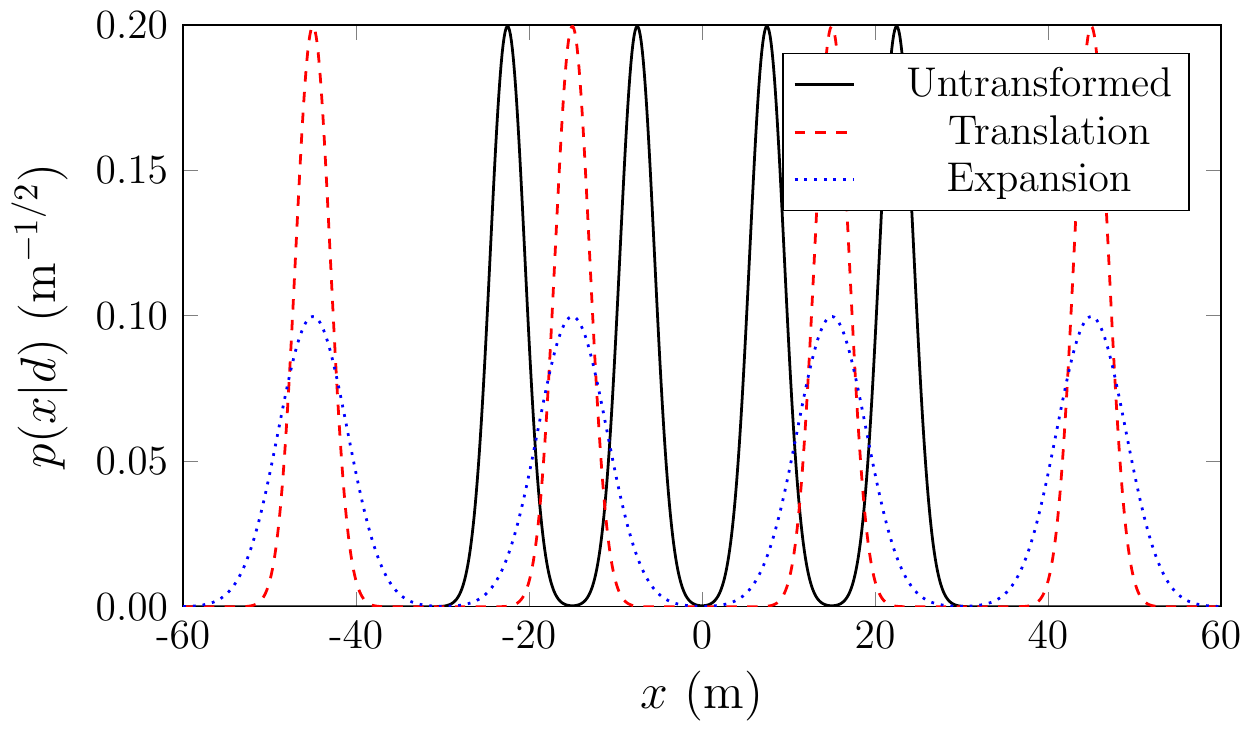}
\caption{(Colour online). Consider the undeformed array of $N=4$ Gaussian spatially profiled sources with a standard deviation of $s=2$ and sources separation distance $d = 15$ shown in solid black. For a simple homogeneous stretching of a 1D array, $\bm{\Xi}_s \rightarrow \xi_s = 2$, the dashed red distribution illustrates the intended behaviour of the generator on the source probability distribution. The expected source positions shift without changing the shape of the probability distribution. The dotted blue distribution represents the unwanted result where the probability distribution is changed implying the nature of the sources changes according the type of transformation considered.}
\label{pic:expansion_translations}
\end{center}
\end{figure}

We consider a 1D array of $N$ identical, stationary and equidistant emitters, each with an intrinsic spatial Gaussian uncertainty $s$. This has been illustrated in Fig.~\ref{pic:stationary_spe_chain} and has been experimentally realised to some extent for near-identical, pure, heralded \textsc{spes}~\cite{Spring2017_O}. We estimate the source separation distance, $d$, after the array is subjected to a general homogeneous deformation. This amounts to a single parameter estimation which would help determine the deformation, $\bm{\Xi}$, and the nature of the sources required to maximise the \textsc{qfi}. Let $\bm{r}$ define the initial coordinates of a source, then after some applied deformation the final source coordinates can be written
\begin{align}
\tilde{\bm{r}} = \bm{\Xi} \cdot \bm{r},
\label{eqn:deformation_matrix}
\end{align}
with the displacement being $\bm{\varepsilon} = (\bm{\Xi} - \mathds{1}) \cdot \bm{r}$. The deformations considered in this work leave the spatial distribution (and hence the variance) of each source invariant and only shifts the expected source positions $\mu_j, \, j \in S_N$ where $S_N$ denotes the set of positive integers $\{1,2, \ldots, N\}$. Cases where the source distribution change would suggest the unlikely scenario where the nature of the sources change with the deformation. Fig.~\ref{pic:expansion_translations} illustrates the differing effects of both types of deformations. Before calculating the \textsc{qfi}, we first derive the generator of translations in the estimating parameter $d$ due to a homogenous deformation.

Without loss of generality, we first consider an array of $N$ \textsc{spe}s, each with a general spatial profile $f(x_j, \mu_j)$ for the $j^{\text{th}}$-source, where $x_j$ defines general coordinates along the array and $\mu_j$ the expected position of source $j$. In what follows, we reserve bold typesetting for tuples. The state may be written
\begin{align}
\begin{split}
\ket{\Psi(d)} &= \bigotimes_{j=1}^N \int \, dx_j \; f(x_j, \mu_j) \hat{a}_j^\dagger(x_j) \ket{0}_j,\\
&= \int \, d\bm{x} \; f(\bm{x}, \bm{\mu}) \hat{a}^\dagger(\bm{x}) \ket{\bm{0}},
\label{eqn:nphoton_unstretched}
\end{split}
\end{align}
where $\smash{f(\bm{x}, \bm{\mu}) = \prod_{j=1}^N f(x_j, \mu_j)}$, $\smash{\hat{a}^\dagger(\bm{x}) = \prod_{j=1}^N \hat{a}_j^\dagger(x_j)}$, $\smash{d\bm{x} = \prod_{j=1}^N dx_j}$ and $\ket{\bm{0}} = \ket{0}^{\otimes N}$ is the multimode vacuum. The expected $j^{\text{th}}$-source position vector is chosen to be symmetric about the array centre $O$ (see Fig.~\ref{pic:stationary_spe_chain}) such that
\begin{align}
\mu_j = \left[j - \frac{(N+1)}{2}\right]d.
\label{eqn:exp_source_positions}
\end{align}
For now we assume each source is mutually independent such that they can be described by separate Hilbert spaces. Hence the mode operators obey the commutation relations $[\hat{a}_j(x), \hat{a}_k^\dagger(y)] = \delta_{jk}\delta(x - y)$ and all other combinations are zero. We later relax this assumption by updating these commutation relations to allow for source overlapping. 


We search for the unitary that generates the new probability distribution after a deformation is applied. Describing the deforming matrix, $\bm{\Xi}$, in 1D as $\xi$, we search for the unitary transformation $\hat{U}(\xi)$ that specifically performs the following operation
\begin{align}
\hat{U}(\xi)\ket{\Psi(d)} = \int \, d\bm{x} \; f(\bm{x}, \tilde{\bm{\mu}}(\xi)) \hat{a}^\dagger(\bm{x}) \ket{\bm{0}},
\label{eqn:nphoton_stretched}
\end{align}
where 
\begin{align}
\tilde{\mu}_j(\xi) = \mu_j + \varepsilon_j(\xi) = \xi \mu_j. 
\label{eqn:transformed_mean_source}
\end{align}
This changes the expected mean positions of the sources, but does not change the source variances. Substituting the state Eq.~(\ref{eqn:nphoton_unstretched}) into Eq.~(\ref{eqn:nphoton_stretched}), Fourier transforming the creation operators and rearranging terms, we find
\begin{align}
\hat{U}(\xi)\hat{a}^\dagger(\bm{k})\hat{U}^\dagger(\xi) = \exp\left[i (\xi-1) \sum_{j=1}^N k_j \mu_j \right] \, \hat{a}^\dagger(\bm{k}).
\label{eqn:stretch_criterion_fourier_domain}
\end{align}
This operation is achieved by the following form of the unitary
\begin{align}
\hat{U}(\xi) = \exp\left[i \, (\xi-1)\sum_{j=1}^N \mu_j \int dk_j \; k_j \, \hat{n}_j(k_j) \right],
\label{eqn:nphoton_unitary}
\end{align}
which may be demonstrated by means of the Baker-Campbell-Hausdorff (\textsc{bch}) identity, and where $\smash{\hat{n}_j(k_j) = \hat{a}_j^\dagger(k_j)\hat{a}_j(k_j)}$ is the number operator of the $j^{\text{th}}$ source in mode $k_j$. Writing $\smash{\hat{U}(\xi) = \bigotimes_{j=1}^N \hat{U}_j(\xi)}$, then $\hat{U}_j(\xi)$ is interpreted as the unitary which performs translations in source $j$ by $\varepsilon_j$. From Eq.~\eqref{eqn:generator_form}, the generator of changes in $d$ due to $\xi$ can be written as
\begin{align}
\smash{\hat{\mathcal{G}}(\xi) = i \hat{U}^\dagger(\xi)\partial_d \hat{U}(\xi)}.
\label{eqn:lie_generator}
\end{align}
Combining with Eq.~\eqref{eqn:nphoton_unitary} provides the final form of our generator
\begin{align}
\hat{\mathcal{G}}(\xi) = - (\xi - 1) \sum_{j=1}^N \mu_j' \int dk_j \; k_j \, \hat{n}_j(k_j),
\label{eqn:nphoton_generator_s_dimensional}
\end{align}
where $\smash{\mu_j' = \partial_d \mu_j = j - (N+1)/2 = \mu_j/d}$. This generator characterises the dynamical property of the parameterisation process of the state on $\bm{\vartheta}$ due to homogenous deformations and its Hermiticity follows from the number operator. It has units of momentum which is expected since the array sources undergo spatial translations according to some homogeneous transformation $\xi$. The unitary in Eq.~(\ref{eqn:nphoton_unitary}) may then be rewritten as
\begin{align}
\hat{U}(\xi) = \exp\left[-i d\hat{\mathcal{G}}(\xi)\right]. 
\label{eqn:general_unitary}
\end{align}
This form implies that the \emph{change} in the \textsc{qfi} as a result of the array deformation may be determined from the variance of the generator~\cite{Braunstein1996_AoP}. 
The \textsc{qfi} of the deformed array is instead computed from the variance of the generator $\smash{\hat{\mathcal{G}}(\xi)}$ with the factor $(\xi - 1)$ in Eq.~\eqref{eqn:nphoton_generator_s_dimensional} replaced by $\xi$. This is a consequence of having derived the unitary by considering the shift in expected source positions, $\smash{\varepsilon_j = (\xi-1) \mu_j}$, resulting from a homogenous deformation, $\xi$. In this representation, the \textsc{qfi} is determined entirely by $\mathcal{\hat{G}}$ and the initial state of the undeformed array. In the next section, we evaluate the variance of the generator for arrays of well defined classical and quantum states of light.


\section{Classical and quantum light sources}
\label{sec:diff_sources}

We consider stretching deformations of the array, with factor $\xi_s$. From the variance of the generator, we calculate the \textsc{qfi} for the array of identical, stationary and equidistant emitters shown in Fig.~\ref{pic:stationary_spe_chain}. We compare the performance of different sources of light on the estimation precision of $d$. This will identify whether difficult to prepare states for enhanced precision measurements is a useful and feasible tradeoff. We choose to work in the near-field regime over the far-field. These define different regions of the electromagnetic field around the sources and allows the use of intensity measurements in the near-field to estimate $d$ as opposed to higher-order correlation measurements in the far-field~\cite{Oppel2012_PRL, Pearce2015_PRA}. For complete state detection, the \textsc{qfi} remains invariant of the regime considered. To demonstrate this, consider complete state detection and the parameter independent unitary, $\hat{U}_{\textsc{ff}}$, which propagates the near-field pure state to the far-field; $\smash{\ket{\Psi_{\textsc{ff}}} = \hat{U}_{\textsc{ff}} \ket{\Psi_{\textsc{nf}}}}$. For a parameter-independent unitary, 
\begin{align}
\ket{\Psi'_{\textsc{ff}}} = \partial_d \ket{\Psi_{\textsc{ff}}} = \hat{U}_{\textsc{ff}} \partial_d\ket{\Psi_{\textsc{nf}}},
\label{eqn:ff_nf_invariance}
\end{align}
we find $\smash{\mathcal{I}^Q_{\textsc{ff}} = \mathcal{I}^Q_{\textsc{nf}}}$. Hence, for complete detection of the state, parameter estimation in both regimes yields the same precision. This equivalence can not be extended to the case of incomplete detection of the state. Such an occurrence may be modelled by considering a near-field calculation comprised of source efficiencies $\eta < 1$. 

\subsection{Single photon emitters}
\label{subsec:spes}

We start by considering $N$ independent \textsc{spe}s. Each is assumed to be generated deterministically with efficiency $\eta =1$. A photon is generated by the $j^{\text{th}}$ source if $\smash{\ket{1}_j = \hat{b}^\dagger_j\ket{0}_j}$, where $\hat{b}_j^\dagger$ is the mode operator describing a photon with a Gaussian spatial profile with centre $\mu_j$ and standard deviation $s$\footnote{We denote the Gaussian standard deviation of the sources as $s$. This distinguishes it from the permutation element $\sigma$ used in Appendix~\ref{sec:appendixA}}. This requires the following form for the creation operator
\begin{align}
\hat{b}_j^\dagger = \frac{1}{(2\pi s^2)^{1/4}} \int dx_j \exp\left[-(x_j - \mu_j)^2/4s^2\right] \hat{a}_j^\dagger(x_j). 
\label{eqn:gaussian_mode_operator}
\end{align}
We initially assume the limit of clear separation, $d \gg s$, where each source may be considered mutually independent. The pure state describing the $j^\text{th}$-source $\smash{\ket{\psi}_j = \hat{b}^\dagger_j\ket{0}_j}$ is then described by its own Hilbert space $\smash{\mathcal{H}_j}$. Defining the total Hilbert space by $\smash{\mathcal{H} = \otimes_{j=1}^N \mathcal{H}_j}$, then the state of the whole array shown in Fig.~\ref{pic:stationary_spe_chain} may be written as the product state
\begin{align}
\ket{\Psi(d)} = \bigotimes_{j=1}^N \ket{\psi}_j
\label{eqn:tot_array_hspace}
\end{align}
where $\ket{\Psi(d)} \in \mathcal{H}$. Since the generator of translations in $d$ is defined in Fourier space, we are required to Fourier transform the state to compute the variance $(\Delta\hat{\mathcal{G}})^2$. Using
\begin{align}
\hat{a}_j^\dagger(x_j) = \frac{1}{\sqrt{2\pi}}\int dk_j \hat{a}_j^\dagger(k_j) \exp\left[ix_jk_j\right]
\label{eqn:mode_operator_ft}
\end{align}
and combining with Eq.~\eqref{eqn:gaussian_mode_operator}, the complete state of $N$ \textsc{spe} in Fourier space may be written as
\begin{align}
\ket{\Psi(d)} = \left(\frac{2 s^2}{\pi}\right)^{\frac{N}{4}} \int \, d\bm{k} \, \exp\left[i \bm{k}\cdot\bm{\mu} - s^2 \bm{k}\cdot\bm{k}\right] \, \hat{a}^\dagger(\bm{k}) \, \ket{\bm{0}},
\label{eqn:pure_state_nf_state}
\end{align}
where the $d$-dependence arises in the tuple $\bm{\mu}$ since $\mu_{j+1} - \mu_j = d$. After an applied stretching of the array about the centre $O$, the \textsc{qfi} of the deformed array of sources is determined from the variance of the 1D generator. %
%
%
%
%
%
The expectation of the generator is zero, which is a consequence of an odd parity integral. Physically, this is since the stretching is performed about the centre of the array and the array does not move as a whole. This is reminiscent of the average momentum of a particle trapped in a harmonic potential well. Hence the \textsc{qfi} is given by $4 \langle\hat{\mathcal{G}}^2(\xi_s)\rangle$. 
\begin{widetext}
\begin{align}
\langle\hat{\mathcal{G}}^2(\xi_s)\rangle = \left(\frac{2 s^2}{\pi}\right)^{\frac{N}{2}}\sum_{j=1}^N \mu_j^2 \int d\bm{k} d\bm{k'} dk''_j dk'''_j k''_j k'''_j \exp\left[i (\bm{k} - \bm{k'}) \cdot \bm{\mu} - s^2 (\bm{k}\cdot\bm{k} + \bm{k'}\cdot \bm{k'})\right] \nu(\bm{k'}, k''_j, k'''_j, \bm{k}),
\label{eqn:generator_squared}
\end{align}
where 
\begin{equation}
\nu(\bm{k'}, k''_j, k'''_j, \bm{k}) = \bra{\bm{0}}\hat{a}(\bm{k'}) \hat{n}_j(k''_j) \hat{n}_j(k'''_j) \hat{a}^\dagger(\bm{k})\ket{\bm{0}}.
\label{eqn:}
\end{equation}
The vacuum expectation value $\nu(\bm{k'}, k''_j, k'''_j, \bm{k})$ for arbitrary $N$ may be written
\begin{equation}
\nu(\bm{k'}, k''_j, k'''_j, \bm{k}) = \delta(k''_j - k'''_j) \delta(k'''_j - k_j) \delta(k'_j - k''_j) \prod_{\substack{i = 1,\\ i\neq j}}^N \delta(k'_i - k_i).
\label{eqn:vev_identity}
\end{equation}
\end{widetext}
Substituting Eq.~(\ref{eqn:vev_identity}) into Eq.~(\ref{eqn:generator_squared}), we obtain the \textsc{qfi} for an array of \textsc{spe}s stretched by factor $\xi_s$ 
\begin{align}
\mathcal{I}^Q_{\textsc{spe}}
 = 4 (\Delta\hat{\mathcal{G}})^2 = \frac{\xi_s^2}{s^2}\sum_{j=1}^N \mu_j'^2 = \frac{\xi_s^2 N (N^2 - 1)}{12 s^{2}}.
\label{eqn:spe_qfi_generator_method}
\end{align}
The first equality holds since the generator is independent of $d$, the second from the variance of the generator. The third equality is from the explicit summation of $\mu_j'$ which was defined earlier. We note that the \textsc{qfi} is independent of the separation distance, $d$, which is welcoming since the parameter to be estimated is often outside the control of the experimenter. A better estimate of $d$ can be achieved by increasing the number of sources $N$ and the stretching factor, $\xi_s$, and decreasing the intrinsic Gaussian emission uncertainty, $s$, as expected. The cubic dependence on $N$ may preliminarily suggest a precision scaling which surpasses the Heisenberg limit. However, the resource count of this physical system is defined by the variance of the generator of translations in $d$~\cite{Giovannetti2006_PRL, Zwierz2012_PRA, Zwierz2010_PRL}, which is not the number of photons. We conjecture that the resource measure is the number of source pairings, which scales quadratically.


If we drop the requirement that $d \gg s$, then neighbouring source distributions may overlap. The resulting \textsc{qfi} is expected to vary with the source separation distance. To understand this dependence, we drop the assumption that each source can be described by a different Hilbert space. This requires different commutations relations to those used earlier, where it was assumed that each source was mutually independent. Specifically, we have $[\hat{a}(x), \hat{a}^\dagger(y)] = \delta(x - y)$ with all other combinations being zero. The Kronecker delta is dropped since we now associate all of the sources with the same Hilbert space $\mathcal{H}$, which can no longer be decomposed in a tensor product structure. These updated commutation relations allow for different source distributions to overlap, and will help determine how the \textsc{qfi} varies with the source separation distance. This demonstration will be made for the array of \textsc{spe} emitters without loss of generality. For distinction with the \textsc{qfi} derived through use of the former commutation relations, we define the \textsc{qfi} calculated with these updated commutations relations as $\mathfrak{I}_\textsc{spe}^Q$.

\begin{figure}[t!]
\begin{center}
\includegraphics[width =\columnwidth]{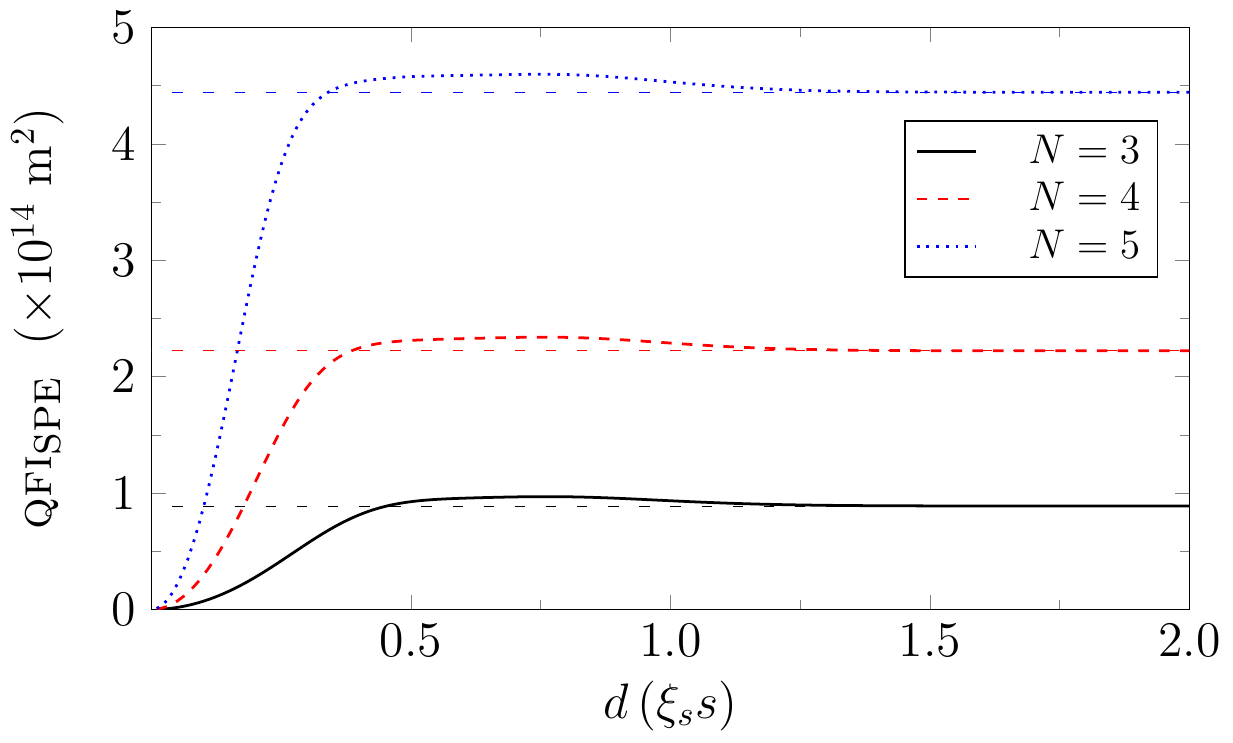}
\caption{(Colour online). The \textsc{qfi} with varying $d$ in units of $s$, where $s = 300$ nm\textemdash typical of photons from quantum dots\textemdash and $\xi_s=2$. Photon bunching is allowed and we note that the \textsc{qfi} approaches the value determined by Eq.~(\ref{eqn:spe_qfi_generator_method}) for $d \sim 2 \xi_ss$. The observed bump preceding the limit of clear separation is a consequence of photon interferences arising due to all the permutations of achieving the same detection.}
\label{pic:qfi_with_varying_d}
\end{center}
\end{figure}

We start by considering a stretched array of $N$ \textsc{spe}s, each emitting photons deterministically with a Gaussian spatial profile. The stretching factor about the array centre is $\xi_s$ and transforms the mean $j^\text{th}$ source position to 
\begin{align}
\tilde{\mu}_j = \left[j - \left(\frac{N+1}{2}\right)\right] \xi_s d.
\label{eqn:trans_mean_pos}
\end{align}
The state describing the $N$ \textsc{spe}s in the near field is written
\begin{align}
\vert\Psi\rangle = \frac{1}{\mathscr{N}} \int_{-\infty}^{\infty} d\bm{x} \; \exp\left[\sum_{j=1}^N\frac{-(x_j - \tilde{\mu}_j)^2}{4s^2}\right] \hat{a}^\dagger(\bm{x}) \, \vert0\rangle,
\label{eqn:field_operator_state}
\end{align}
where $\hat{a}^\dagger(\bm{x}) = \hat{a}^\dagger(x_1) \cdots \hat{a}^\dagger(x_N)$ and $\mathscr{N}$ is the normalisation constant. We note the subtle change to the notation used for the vacuum state and the mode operators which now span the entire Hilbert space $\mathcal{H}$. This is in contrast to Eq.~\eqref{eqn:pure_state_nf_state} where mode operators corresponding to the $j^\text{th}$-source acted only on its associated Hilbert space. Defining
\begin{align}
g_{\alpha}^\beta = \exp\left[-\frac{1}{2s^2} \sum_{k = \alpha}^\beta x_k^2 - x_k(\tilde{\mu}_k + \tilde{\mu}_{\sigma(k)})\right],
\label{eqn:functional_form_g}
\end{align}
then
\begin{align}
\vert\mathscr{N}\vert^2 = \exp\left[\frac{-d^2\mathcal{I}^Q_\textsc{spe}}{2}\right]\sum_\sigma \int^{\infty}_{-\infty} d \bm{x} \, g_{1}^N
\label{eqn:normalisation_field_operator}
\end{align}
where we recall the definition of $\mathcal{I}^Q_\textsc{spe}$ from Eq.~\eqref{eqn:spe_qfi_generator_method} in the main section and $\sigma$ denotes all of the possible permutations associated with the number of sources, $N$. Since Eq.~\eqref{eqn:field_operator_state} is a pure state the \textsc{qfi} may be determined from 
\begin{align}
\mathfrak{I}^Q_\textsc{spe} = 4 \left\{\langle \Psi' \vert \Psi' \rangle - \left| \langle\Psi \vert\Psi'\rangle \right \vert ^2 \right \},
\label{eqn:field_operator_qfi_final_result}
\end{align}
where $\smash{\vert\Psi'\rangle = \partial_d\vert\Psi\rangle}$. We later realise this method provides the same result for the \textsc{qfi} as that determined from the variance of the generator $\hat{\mathcal{G}}(\xi_s)$, providing a convincing verification. Noting that the normalisation constant has no dependence on the integration variable, 
then by use of the following vacuum expectation value
\begin{align}
\langle0\vert\prod_{j=1}^N \hat{b}(x_j)\prod_{j=1}^N \hat{b}^\dagger(x'_j)\vert0\rangle = \sum_\sigma\prod_{j=1}^N \delta\left(x_j - x'_{\sigma(j)}\right),
\label{eqn:identity1}
\end{align}
and the permutation group identities summarised in Appendix~\ref{sec:appendixA}, we find
\begin{align}
\begin{split}
\langle\Psi\vert\Psi'\rangle &= \gamma + \frac{e^{-d^2\mathcal{I}^Q_\textsc{spe}/2}}{2s^2\vert\mathscr{N}\vert^2} \sum_\sigma \int_{-\infty}^{\infty} d\bm{x} \left(\sum_{j=1}^N \tilde{\mu}_j' x_j - ds^2\mathcal{I}^Q_\textsc{spe}\right) g_1^N, \\
&= \gamma - \frac{d\mathcal{I}^Q_\textsc{spe}}{2} + \frac{e^{-d^2\mathcal{I}^Q_\textsc{spe}/2}}{2s^2\vert\mathscr{N}\vert^2} \sum_\sigma \left[\sum_{j=1}^N \tilde{\mu}_j' \int_{-\infty}^{\infty} d\bm{x} \, x_j \, g_1^N\right],
\label{eqn:fi_term2_field_operator_2}
\end{split}
\end{align}
where we defined the constant $\gamma = \gamma(d, N) = \partial_d\left[\ln\left(\frac{1}{\mathscr{N}}\right)\right]$. The second line used the definition of the normalisation constant Eq.~\eqref{eqn:normalisation_field_operator}. To ease the notation, we denote the last term in Eq.~\eqref{eqn:fi_term2_field_operator_2} as $\mathscr{B}$. Similarly,
\begin{widetext}
\begin{align}
\langle\Psi'\vert\Psi'\rangle = \gamma\left(\gamma + 2\mathscr{B} - d\mathcal{I}^Q_\textsc{spe}\right) + \frac{d^2(\mathcal{I}^Q_\textsc{spe})^2}{4} - \mathscr{B}d\mathcal{I}^Q_\textsc{spe} + \frac{e^{-d^2\mathcal{I}^Q_\textsc{spe}/2}}{4s^4\vert\mathscr{N}\vert^2} \sum_\sigma \left[\sum_{j, k =1}^N  \tilde{\mu}_{\sigma(j)}' \tilde{\mu}_k' \int_{-\infty}^{\infty} d\bm{x} \, x_j x_k \, g_1^N\right].
\label{eqn:fi_term1_field_operator}
\end{align}
\end{widetext}
We define $\mathscr{C}$ to be the last term in Eq.~\eqref{eqn:fi_term1_field_operator}. From Eq.~\eqref{eqn:field_operator_qfi_final_result} we get the following for the \textsc{qfi}
\begin{align}
\mathfrak{I}^Q_\textsc{spe} = 4 \left(\mathscr{C} - \left\vert\mathscr{B}\right\vert^2\right).
\label{eqn:field_operator_qfi_final_result_2}
\end{align}
Despite this simplicity, the evaluation of the \textsc{qfi} for particular values of $d$ can only be addressed through a numerical approach due to the sum over all permutations associated with the two expressions involved. It has no dependence on $\gamma$, which is expected since the normalisation constant trivially has no physical contribution to the information in the system. On the numerical front, the simplicity of Eq.~\eqref{eqn:field_operator_qfi_final_result_2} provides a two-fold advantage. Firstly, there are fewer terms to evaluate. Secondly, this term dominated the value of all other terms which contributes the \textsc{qfi}. This domination saw the resulting difference between $\langle \Psi'\vert\Psi'\rangle$ and $\langle \Psi'\vert \Psi\rangle$ be zero. A solution to overcome this would be to increase the working precision of the numerical analysis at the expense of greater computational time. Identifying this cancellation proves this unnecessary. We note that the evaluation of the \textsc{qfi} is reduced to that of two terms only; $\mathscr{B}$ and $\mathscr{C}$. However, both terms contain multi-dimensional integrals over all possible permutations for any given $N$. The computation time to evaluate this using a brute-force method increases rapidly rendering this unsuitable. We address this by taking a functional-approach to the problem. This is possible since both terms $\mathscr{B}$ and $\mathscr{C}$ are comprised of repeat integrals, differing only in the index of the source positions. For convenience, both terms are written here
\begin{align}
\mathscr{B} &= \frac{e^{-d^2\mathcal{I}^Q_\textsc{spe}/2}}{2s^2\vert\mathscr{N}\vert^2} \sum_\sigma \left(\sum_{j=1}^N  \tilde{\mu}_j' \int_{-\infty}^{\infty} d\bm{x} \, x_j \, g_1^N\right), \\
\mathscr{C} &= \frac{e^{-d^2\mathcal{I}^Q_\textsc{spe}/2}}{4s^4\vert\mathscr{N}\vert^2} \sum_\sigma \left(\sum_{j, k =1}^N  \tilde{\mu}_{\sigma(j)}'  \tilde{\mu}_k'\int_{-\infty}^{\infty} d\bm{x} \, x_j x_k \, g_1^N\right).
\label{eqn:terms2_defs}
\end{align}
Both contain repeating Gaussian integrals of the type $h_n = \int^{\infty}_{-\infty} dx_j \; x_j^n g_j^j$. By analytically solving the necessary terms
\begin{align} 
h_0 &= \sqrt{2 \pi  s^2} \exp\left[\frac{(\tilde{\mu}_j + \tilde{\mu}_{\sigma(j)})^2}{8 s^2}\right], \nonumber \\
h_1 &= \sqrt{\frac{\pi  s^2}{2}} \left(\tilde{\mu}_j + \tilde{\mu}_{\sigma(j)}\right) \exp\left[\frac{(\tilde{\mu}_j + \tilde{\mu}_{\sigma(j)})^2}{8 s^2}\right], \\ 
h_2 &= \sqrt{\frac{\pi  s^2}{8}} \left[\left(\tilde{\mu}_j + \tilde{\mu}_{\sigma(j)}\right)^2+4 s^2\right] \exp\left[\frac{(\tilde{\mu}_j + \tilde{\mu}_{\sigma(j)})^2}{8 s^2}\right]. \nonumber 
\end{align}
we can rewrite the terms. Simplifying, we find that
\begin{widetext}
\begin{subequations}
\begin{equation}
\mathscr{B} = \frac{\sum_\sigma \left\{\sum_{j=1}^N \left[\tilde{\mu}_j' (\tilde{\mu}_j + \tilde{\mu}_{\sigma(j)})\right] \exp\left[\sum_{k=1}^N \frac{\tilde{\mu}_k\tilde{\mu}_{\sigma(k)}}{4s^2}\right]\right\}}{4s^2 \sum_\sigma \exp\left[\sum_{l=1}^N \frac{\tilde{\mu}_l\tilde{\mu}_{\sigma(l)}}{4s^2}\right]}, 
\label{eqn:near_field_terma_functional_form}
\end{equation}
\begin{equation}
\mathscr{C} = \frac{\sum_\sigma\left(\left\{ \sum_{j=1}^N \tilde{\mu}_j'\tilde{\mu}_{\sigma(j)}' \left[(\tilde{\mu}_j + \tilde{\mu}_{\sigma(j)})^2 + 4s^2\right] + \sum_{\substack{j, k = 1,\\j \neq k}} \tilde{\mu}_{\sigma(j)}'\tilde{\mu}_k' (\tilde{\mu}_j + \tilde{\mu}_{\sigma(j)}) (\tilde{\mu}_k + \tilde{\mu}_{\sigma(k)}) \right\}\exp\left[\sum_{l=1}^N \frac{\tilde{\mu}_l\tilde{\mu}_{\sigma(l)}}{4s^2}\right]\right)}{16s^4 \sum_\sigma \exp\left[\sum_{l=1}^N \frac{\tilde{\mu}_l\tilde{\mu}_{\sigma(l)}}{4s^2}\right]}.
\label{eqn:near_field_termb_functional_form}
\end{equation}
\label{eqn:near_field_functional_forms}
\end{subequations}
\end{widetext}
The same result is reached if the calculation were repeated using instead the variance of the generator. We find that the \textsc{qfi} from Eq.~\eqref{eqn:field_operator_qfi_final_result_2} depends only on the properties of the source positions. This is a property of the \textsc{qfi} which depends only on the state. Since it contains a sum over all possible permutations of source overlaps, a reduced analytic form is not possible and a numerical approach is taken. The simulation results illustrated in Fig.~\ref{pic:qfi_with_varying_d} indeed show a dependence with $d$. As expected, for small separation distances the estimation precision increases with increasing $d$. For larger source separations where neighbouring sources are spatially distinct, the \textsc{qfi} converges to the value governed by Eq.~(\ref{eqn:spe_qfi_generator_method}). This mutual independence was assumed to hold for $d \gg s$. However, we observe from Fig.~\ref{pic:qfi_with_varying_d} that this regime is in fact satisfied when $d \geq 2s \; \forall \; N$ for a unit (untransformed) stretching factor. 
Between these two regimes, a small bump is observed between $d \sim (0.2 -1.5)s$. 
This nearest-neighbour effect persists for arbitrarily large number of sources, $N$, on the array. The observed protuberance in Fig.~\ref{pic:qfi_with_varying_d} appears to suggest that light sources with a higher average mode occupancy may be preferential for the estimation of $d$. To test this, we will next determine the precision scalings achievable with an array of coherent and thermal states. If this proposition is found to be true, the use of coherent and thermal states over \textsc{spe}s in this context would give a better estimate of $d$.
%
In what follows, we shall assume each source to be mutually independent.

%

\subsection{Coherent sources}
\label{subsec:coherent}
The semi-classical single mode coherent state is defined as
\begin{align}
\vert\alpha_j\rangle = \exp\left[\frac{-\vert\alpha_j\vert^2}{2}\right]\sum_{n} \frac{\alpha_j^n \hat{b}_j^{\dagger n}}{n!} \vert0\rangle,
\label{eqn:coherent_state}
\end{align}
where $\alpha_j = r_j\exp[i\varphi_j]$ is the amplitude associated with the source mode $j \in S_N$ and $b_j^\dagger$ is the creation operator for that mode. To encode the source separation distance $d$ into Eq.~(\ref{eqn:coherent_state}), we use the same mode creation operator $\hat{b}_j = \int dx_j f(x_j, \mu_j) \hat{a}_j^\dagger(x)$ as defined for the \textsc{spe}s, where the function $f(x_j, \mu_j)$ defines the spatial Gaussian profile. Assuming negligible overlap between different sources, the $N$--mode coherent state, $\smash{\vert\Psi\rangle_\mathrm{c} = \otimes_{j=1}^N \vert\alpha_j\rangle}$, is written 
\begin{widetext}
\begin{align}
\begin{split}
\vert\Psi\rangle_\mathrm{c} &= \exp\left[-\frac{N \vert\alpha\vert^2}{2}\right] \frac{1}{\mathscr{N}_\text{c}}\displaystyle \sum_{n_1 = 0}^\infty \cdots \sum_{n_N = 0}^\infty \left[\prod_{j=1}^N \frac{\left(r\exp[i\varphi_j]\int dx_j \; f(x_j) \; \hat{a}_j^\dagger(x_j)\right)^{n_j}}{n_j!}\right]\; \vert\bm{0}\rangle,\\
&= \exp\left[-\frac{N \vert\alpha\vert^2}{2}\right] \frac{1}{\mathscr{N}_\text{c}} \displaystyle \sum_{\{\bm{n}\} \geq 0} \mathscr{A}(\bm{n}) \int d\bm{x}^{\bm{n}} \; f(\bm{x})^{\bm{n}} \; \hat{a}^{\bm{\dagger} \bm{n}}(\bm{x}) \; \vert\bm{0}\rangle,
\end{split}
\label{eqn:n_coherent_state}
\end{align}
where $\bm{n}! = n_1! n_2! \cdots n_N!$, $\smash{\bm{\alpha}(\bm{x})^{\bm{n}} = \prod_{j=1}^N\alpha_j(x_j)^{n_j}}$, $\smash{\hat{a}^{\bm{\dagger} \bm{n}}(\bm{x}) = \prod_{j=1}^N \hat{a}_j^{{\dagger} {n_j}}(x_j)}$, $\mathscr{N}_\text{c} =\,_\text{c}\hspace{-0.005cm}\braket{\Psi\vert\Psi}_\text{c}$ is the normalisation constant and $\mathscr{A}(\bm{n}) = \prod_{j=1}^Nr^{n_j}\exp[i \varphi_j n_j]/n_j!.$
\bigskip
\end{widetext}
Applying a stretching with factor $\xi_s$, we can calculate the \textsc{qfi} of the new state from variance of the generator yielding
\begin{align}
\mathcal{I}^Q_{\text{Coherent}}  = \frac{\xi_s^2 \exp[-N r^2]}{s^2} \sum_{\{\bm{n}\} \geq 0} \prod_{j=1}^N\left(\frac{r^{2n_j}}{n_j!}\right) \sum_{k = 1}^N \mu_k'^2 n_k^2.
\label{eqn:coherent_nf_qfi}
\end{align}
We immediately observe that the \textsc{qfi} is independent of the phase $\varphi_j$ of the coherent states. Although the \textsc{qfi} is a function of the state alone, this independence may be understood by considering the phase difference of any two sources, $\Delta\varphi$, at any region of space along the near field plane. A change to the phase difference $\Delta\varphi \mapsto \Delta\varphi + \varphi_0$ only occures if any one of the two sources contributing the phase difference shifts along the array. However, since
\begin{align}
\frac{\partial}{\partial x}(\Delta\varphi + \varphi_0) = \frac{\partial \Delta\varphi}{\partial x},
\label{eqn:phase_diff_indep}
\end{align}
any changes to the phase of each source do not contribute to the overall \textsc{qfi}. Any information change is encoded in the separation distance.

A meaningful comparison with the \textsc{spe}s requires a unit average photon number in the $N$-coherent state, such that $\Braket{\hat{n}} = \Abs{\alpha}^2 = r^2 = 1$. For the limiting value $n_j \rightarrow \infty \; \forall \; j \in S_N$, Eq.~(\ref{eqn:coherent_nf_qfi}) takes a similar form to $\smash{\mathcal{I}^Q_{\textsc{spe}}}$ as follows
\begin{align}
\mathcal{I}^Q_{\text{Coherent}}  = \frac{\xi_s^2 N(N^2 - 1)}{6 s^2}.
\label{eqn:coherent_nf_qfi_2}
\end{align}
We note that the scaling with resources is similar to that of the \textsc{spes}, and the constant factor of 2 improvement results from the increased mode occupancy of the coherent states.


\subsection{Thermal sources}
\label{subsec:thermal_states}

Another class of widely occurring states in nature are the thermal states. In this section, we replace the array of \textsc{spe} sources for the classical thermal states. This would address a comparative performance on the estimation on the source separation distance $d$. A thermal state emits at all frequencies with an intensity determined by the Planck distribution. This distribution can be considered an infinite number of independent spectral modes~\cite{Barnett2005, Walls2008}. The $N$-mode blackbody distribution is defined
\begin{align}
\rho_{\text{Bb}} = \bigotimes_{j = 1}^N \rho_\text{Th}^j = \sum_{\{\bm{n}\} \geq 0} \; \frac{c_{\bm{n}}}{\bm{n!}} \; \hat{b}^{\dagger \bm{n}}_{\bm{k}}\vert \bm{0}\rangle\langle \bm{0}\vert \hat{b}^{\bm{n}}_{\bm{k}},
\label{eqn:multimode_thermal_state}
\end{align}
where the total spectral mode creation operator
\begin{align}
\smash{\hat{b}^{\dagger \bm{n}}_{\bm{k}} = \bigotimes_{j = 1}^N \hat{b}_{k_j}^{\dagger n_j}}
\label{eqn:spectral_mode_cr_op}
\end{align}
is composed from the tensor product over the $j^{\text{th}}$ source mode operator and $\smash{\bm{n!} = \prod_{j=1}^N n_j!}$. The photon-counting distribution or occupancy number, $c_{\bm{n}}$ is determined from the Bose-Einstein probability distribution and has the form
\begin{align}
c_{\bm{n}} = \prod_{j=1}^N\frac{\overline{n}_j^{n_j}}{(1 + \overline{n}_j)^{1 + n_j}},
\label{eqn:therma_occupancy_number}
\end{align}
where
\begin{align}
\overline{n}_j = \Braket{\hat{n}_j} = \Tr{\rho_\text{Bb} \hat{n}_j}
\label{eqn:ave_therm_photon_num}
\end{align}
is the mean photon number for the $j^\text{th}$-source. The creation operator $\smash{\hat{b}_j^\dagger}$ defines the same inherent Gaussian uncertainty in the position basis as that used for the coherent states. Since the blackbody distribution is defined in the Fourier space, we are required to use the Fourier transform of the field operator in Eq.~\eqref{eqn:gaussian_mode_operator} for a single mode. This yields
\begin{align}
\hat{b}^\dagger_{j} = \left(\frac{2s^2}{\pi}\right)^{1/4} \int dk_j \; \hat{a}_j^\dagger (k_j) \exp\left[k_j (-k_js^2 + i \mu_j)\right],
\label{eqn:fourier_b_operator}
\end{align}
which upon substitution into Eq.~(\ref{eqn:multimode_thermal_state}) yields the final form of the array of thermal states $\rho_{\text{Bb}}$.
%
%
It describes thermal states produced at positions $\mu_j$, each with an average number of photons $\overline{n}_j$. It runs along the continuous variable $k$, in difference to Eq.~(\ref{eqn:multimode_thermal_state}) which describes a discrete combination over the different Hilbert spaces associated with each source.

The \textsc{qfi} for multimode states becomes additive such that for the blackbody state in Eq.~\eqref{eqn:multimode_thermal_state} we can write
\begin{align}
\mathcal{I}^Q_\text{Bb} = \sum_{j=1}^N \left[\mathcal{I}^Q_\text{Th}\right]_j,
\label{eqn:qfi_blackbody}
\end{align}
where the sum is over the thermal modes. Since the thermal states are mixed states, the variance of the generator only provides an upper bound to the \textsc{qfi}. Hence, using the full form of the \textsc{qfi} in Eq.~\eqref{eqn:qfi_generator_form} for an arbitrary thermal state we have  
\begin{align}
\mathcal{I}^Q_\text{Th} = 4 \sum_{k=0}^\infty p_k \left(\Delta \mathcal{\hat{G}}^2\right)_k - \sum_{\substack{{k, l =0,} \\ {k \neq l}}}^\infty \frac{8p_kp_l}{p_k + p_l} \Abs{\Braket{\varrho_k \left\vert \mathcal{\hat{G}} \right\vert \varrho_l}}^2
\label{eqn:qfi_thermal_mode}
\end{align}
where the probabilities and eigenstates of the $j^\text{th}$-mode thermal state $\rho_\text{Th}^j$ are given by
\begin{align}
p_k = \frac{c_{n_k}}{n_k!}, \qquad \ket{\varrho_k}_j = \hat{b}^{\dagger n_k}_j \ket{0}_j.
\label{eqn:thermal_evalues_vectors}
\end{align}
As expected, we find that the expectation of the generator is zero since stretching the array does not move the photons. We find
\begin{align}
\begin{split}
\mathcal{I}^Q_\text{Bb} &= \frac{\xi_s^2}{s^2}\sum_{j=1}^N \mu_j'^2 \sum_{n_j=0}^\infty \frac{n_j^2 \overline{n}_j^{n_j}}{(1 + \overline{n}_j)^{1 + n_j}}, \\
&= \frac{\xi_s^2}{s^2}\sum_{j=1}^N \mu_j'^2 \overline{n}_j (1 + 2\overline{n}_j),
\end{split}
\label{eqn:qfi_blackbody_2}
\end{align}
where the second equality made use of the infinite summation identity $\sum_k k^2 a^k = a(1+a)/(1-a)^3$ since $\abs{\overline{n}_j/1+\overline{n}_j} < 1$. For a meaningful comparison with the \textsc{spe}s we take a unit average photon number, requiring $\braket{\hat{n}_j} = \overline{n}_j = 1$. This gives
\begin{align}
\mathcal{I}^Q_{\text{Bb}}  = \frac{\xi_s^2N(N^2 - 1)}{4 s^2}.
\label{eqn:qfi_blackbody_3}
\end{align}
%
%
\begin{figure}[t!]
\begin{center}
\includegraphics[width =\columnwidth]{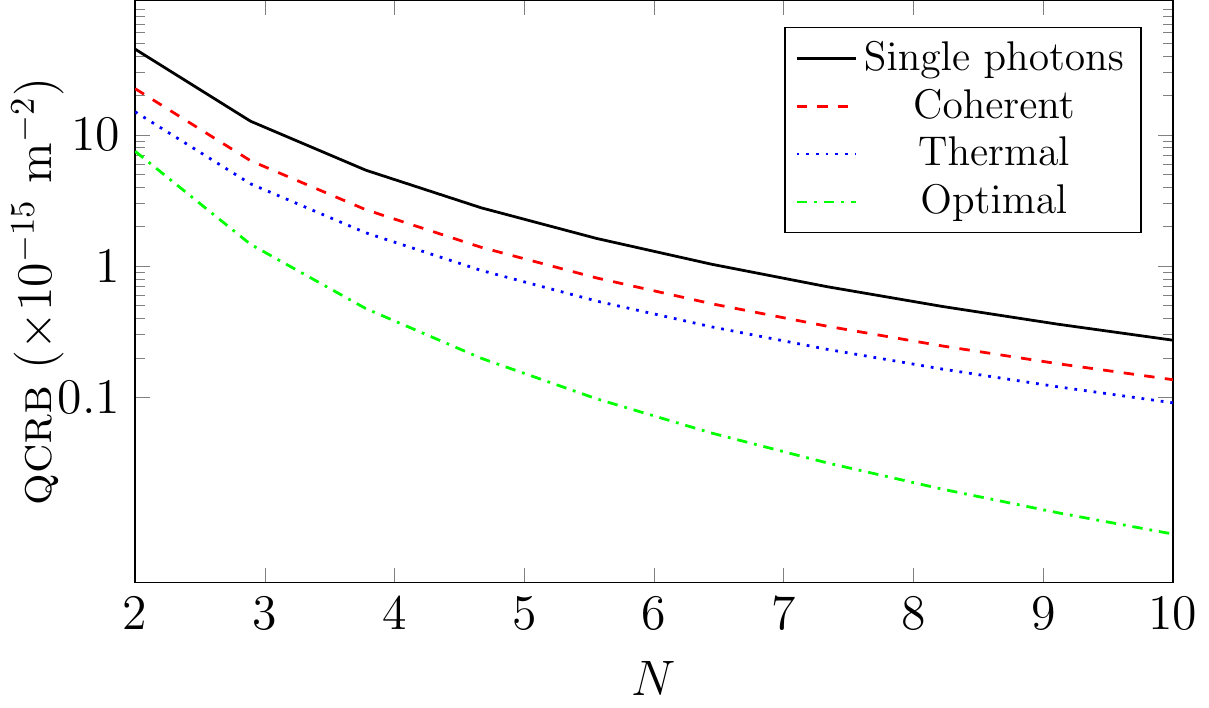}
\caption{(Colour online). \textsc{qcrb} scaling with $N$ for \textsc{spe}s, coherent, thermal states and the optimally entangled state Eq.~\eqref{eqn:optimal_entangled_state} with $s = 300$ nm and $\xi_s=2$. The higher mode occupancy of thermal states permits better estimation performance when compared with the quantum \textsc{spe}s. However, the optimal state remains the entangled state constructed from the eigenvectors corresponding to the minimum and maximum eigenvalues of the generator $\hat{\mathcal{G}}$.}
\label{pic:qcrb_different_sources}
\end{center}
\end{figure}%
%
%
Fig.~\ref{pic:qcrb_different_sources} illustrates this scaling with the number of thermal sources $N$ along the array. Earlier studies have shown that with data post-processing, higher order correlations of \textsc{spe}s yield more information than thermal light sources~\cite{Oppel2012_PRL}. Here, we find that thermal states provide a better estimate of the source separation distance for a single shot experiment ($\nu=1$) in the absence of any post-processing techniques.

\subsection{Entangled states of \textsc{spe}s and the optimal state}
\label{subsec:entangled_spes}

In this section we consider how entanglement can be used as a resource to improve the estimation precision for detection of spatial deformations. Since the \textsc{qfi} is a property of the quantum state alone and does not depend on a particular measurement scheme, the estimation precision is limited only by the uncertainty in the state only. On the theoretical front, the optimal state defines the statistical properties of the probe state which saturates the \textsc{qcrb}. It does not address the optimal measurement strategy that should be employed. We defer a discussion of this to the next section. 

Giovannetti \emph{et al.} showed the optimal state to be that which entangles the states corresponding to the maximum and minimum eigenvectors of the generator $\mathcal{\hat{G}}$~\cite{Giovannetti2011_NP}. Specifically, we consider 
\begin{align}
\ket{\Psi}_\mathrm{Opt} = \frac{\ket{\psi}_\mathrm{max} + \ket{\psi}_\mathrm{min}}{\sqrt{2}},
\label{eqn:optimal_entangled_state}
\end{align}
where $\ket{\psi}_\mathrm{max}$ is the state corresponding to the maximum eigenvalue of the generator such that $\smash{\mathcal{\hat{G}} \ket{\psi}_\mathrm{max} = g_\mathrm{max}\ket{\psi}_\mathrm{max}}$ and similarly for the state $\ket{\psi}_\mathrm{min}$. Then, from the variance of the generator we have
\begin{align}
\mathcal{I}^Q_\mathrm{Opt} = 4\Delta\mathcal{\hat{G}}^2 = \left(g_\mathrm{max} - g_\mathrm{min}\right)^2.
\label{eqn:qfi_optimal_state}
\end{align}
From the matrix elements of the generator in the momentum Fock-basis 
\begin{align}
\braket{n_\alpha(k_\alpha)\vert\mathcal{\hat{G}}\vert n_\beta(k_\beta)} = \xi_s \mu_\beta'k_\alpha n_\beta(k_\beta) \delta_{\alpha \beta}
\label{eqn:g_matrix_elements}
\end{align}
where $\alpha, \beta \in S_N$, we see that the generator is diagonal with eigenvalues given when $\alpha = \beta$. Given the definition of the source positions in Eq.~\eqref{eqn:exp_source_positions}, the maximum eigenvalue corresponds to $\alpha=N$ and the minimum for $\alpha=1$ when $n_1(k_1) = n_N(k_N) = N$. 
This finding could have been consistently anticipated from the results obtained in the preceding subsections. For the same resource count, all of the calculations for the \textsc{qfi} for the different sources considered contained the term $\sum_j \mu_j'^2$. Hence the optimal state-that which maximises the \textsc{qfi}-would have all $N$-photons emitting from the most extremal positions about the array centre $O$. From this, we construct the optimally entangled state in Eq~\eqref{eqn:optimal_entangled_state} by identifying
\begin{align}
\begin{split}
\ket{\psi}_\mathrm{\text{max}} & = \int dx^N f(x, \mu_{N})^N \frac{\hat{a}^\dagger_{N}(x)^N}{\sqrt{N!}} \ket{\bm{0}} \\
\ket{\psi}_\mathrm{\text{min}} & = \int dx^N f(x, \mu_{1})^N \frac{\hat{a}^\dagger_{1}(x)^N}{\sqrt{N!}} \ket{\bm{0}}
\end{split}
\label{eqn:optimal_max_min_state}
\end{align}
Then from the variance of the generator we obtain
\begin{align}
\mathcal{I}^Q_\mathrm{Opt} = \frac{\xi_s^2 N^2 (N-1)^2}{4s^2}.
\label{eqn:optimal_qfi}
\end{align}
The scaling of the \textsc{qcrb} with $N$ has been illustrated in Fig.~\ref{pic:qcrb_different_sources}. As expected, it outperforms the \textsc{spe}s, coherent and thermal states. 
We note that not all entangled states reproduce a better performance than the classical states. To demonstrate this, consider the following simple entangled state of \textsc{spe}s 
\begin{align}
\ket{\psi(d)} = \sqrt{p} \ket{\psi(d)}_\text{odd} + \sqrt{1 - p} \ket{\psi(d)}_\text{even},
\label{eqn:spe_entangled_state}
\end{align}
which emits single photons from either the odd sources along the array, $\smash{\ket{\psi(d)}_\text{odd}}$, or the even sources, $\ket{\psi(d)}_\text{even}$. From the variance of the generator we obtain 
\begin{align}
\mathcal{I}^Q_{\textsc{e-spe}} = \frac{\xi_s^2}{s^2}\left[p\sum_{\substack{j=1, \\ j = \text{odd}}}^{N} \mu_j'^2 + (1 - p) \sum_{\substack{j=2, \\ j = \text{even}}}^N \mu_j'^2\right],
\label{eqn:spe_entangled_qfi_generator_method}
\end{align}
%
%
which for a maximally entangled state $p = 1/2$ reduces to
\begin{align}
\mathcal{I}^Q_{\textsc{e-spe}} = \frac{1}{2}\mathcal{I}^Q_{\textsc{spe}}.
\label{eqn:QFI_entangled_comparison}
\end{align}
Hence entanglement as a resource does not necessarily always provide precision enhancements. This is reminiscent of separate studies in both optical imaging and quantum computing where entanglement was necessary but insufficient in providing performance enhancements~\cite{Tsang2008_PRL, Gottesman1997_Thesis}. Whilst the entangled state introduced by Giovannetti \emph{et al.} remains optimal, it is constructive to acknowledge the increasing number of researchers who consider entanglement unnecessary to achieve resolutions beyond the diffraction limit~\cite{Tilma2010_PRA, Braun2017_arxiv}



\section{Optimal estimator}
\label{sec:optimal_states}

An optimal estimator is one that saturates the \textsc{qcrb}. From Eq.~\eqref{eqn:optimal_measure_operator} we observe that it corresponds to the source that minimises the covariance of the estimator. We first consider intensity measurements. If found to be optimal, we expect the classical Fisher information (\textsc{cfi}) for photon number counting to become identical to its corresponding \textsc{qfi}.

The \textsc{cfi} for the stretched array of $N$ independent \textsc{spe}s with $\eta=1$. The detector is placed in the near field and is discretised into $M$ pixels which covers the entire spatial extent of the array. For some state $\rho(d)$ incident on the detector, the measurement is generally described by a positive operator-valued measure \textsc{povm}. Intensity measurements of the state are most common in imaging and can often be described by operators which are diagonal in the Fock basis. Hence we write the probability distribution of number counting at each pixel as
\begin{align}
p(n_1, \ldots, n_M) & = \Tr{\rho(d)\ket{n_1}\bra{n_1} \otimes \cdots \otimes\ket{n_M}\bra{n_M}}, \nonumber \\
& = \Tr{\rho(d)\ket{\vec{n}}\bra{\vec{n}}}, \label{eqn:prob_distribution_intensities} \\
& = \left\vert\braket{\vec{n} \vert \psi(d)}\right\vert^2, \nonumber
\end{align}
where we use the notation $\vec{n}$ is used to define vectors spanning Hilbert space of the detector. We note that the form of the probability distribution in Eq.~\eqref{eqn:prob_distribution_intensities} is motivated by photon counting and the separable form of the \textsc{spe} state in Eq.~\eqref{eqn:tot_array_hspace}. Since we assume the sources to be well separated, then in the near field it is unlikely that more than one photon is detected at the same pixel. This truncates the Fock basis of each of the sources to values in the set $n_k = \{0,1\} \; \forall \; k \in S_N$. Hence we find that 
\begin{align}
p(\vec{n}\vert d) = \prod_{j=1}^N \left\vert f(x_j, \tilde{\mu}_j)\right\vert^2,
\label{eqn:photon_counting_prob_distb}
\end{align}
where $f(x_j, \tilde{\mu}_j)$ defines a normalised Gaussian centred on $\tilde{\mu}_j$ and standard deviation $s$. We then obtain 
\begin{align}
\mathcal{I}^C_{\textsc{spe}} = \int d\bm{x} \frac{1}{p(\vec{n}\vert d)}\left(\frac{\partial p(\vec{n}\vert d)}{\partial d}\right)^2 = \frac{\xi_s^2N (N^2 - 1)}{12 s^2},
\label{en:classical_fi}
\end{align}
for the near field \textsc{cfi}. We find that the \textsc{cfi} is equal to the \textsc{qfi}, which is since the probability distribution in Eq.~\eqref{eqn:photon_counting_prob_distb} is the same as that describing the state $\rho(d)$ of the array of \textsc{spe}s. As intuition may suggest, this equivalence implies that photon-number counting in the near field is the optimal measurement strategy which saturates the \textsc{qcrb}. To examine this statement further we note that from Eq.~(\ref{eqn:optimal_measure_operator}), the optimal estimator is given by the eigenbasis of the \textsc{sld} for single parameter estimations. Since for pure states $\smash{\rho(d) = \rho(d)^2}$, then from the implicit definition of the \textsc{sld} in Eq.~\eqref{eqn:sld_implicit_definition}, we have~\cite{Fujiwara1995_PRA}
\begin{align}
\hat{\mathscr{L}}(d) = 2\partial_d\rho(d).
\label{eqn:pure_state_sld}
\end{align}
The optimal estimator then becomes
\begin{widetext}
\begin{align}
\check{O}(d) = d \mathds{1} + \frac{12}{\xi_s^2 N (N^2 - 1)} \sum_{j=1}^N \mu_j' \int_{-\infty}^{\infty} d\bm{x} d\bm{x'} f(\bm{x}) f(\bm{x'}) \left(x_j + x_j' - 2\mu_j\right)\hat{a}^\dagger(\bm{x})\Ket{\bm{0}}\Bra{\bm{0}}\hat{a}(\bm{x'}).
\label{eqn:optimal_estimator}
\end{align}
\bigskip
\end{widetext}
To check the optimality of the estimator we confirm its variance reproduces the inverse of the \textsc{qfi} in Eq.~\eqref{eqn:spe_qfi_generator_method} for a single shot experiment $\nu = 1$. For shorthand, we redefine Eq.~\eqref{eqn:optimal_estimator} as $\smash{\check{O}(d) = d\mathds{1} + \hat{Q}}$ where the first term ensures that the estimator is unbiased since $\tr{\check{O}(d) \rho(d)} = d$. From
\begin{align}
\check{O}(d)^2 = d (\check{O}(d) + \hat{Q}) + \hat{Q}^2,
\label{eqn:estimator_sqr_exptn}
\end{align}
and since the expectation of the \textsc{sld} is zero, we have
\begin{align}
\braket{\check{O}(d)^2} = d^2 + \Braket{\hat{Q}^2}.
\label{eqn:estimator_sqr_exptn}
\end{align}
From this we find the characteristic condition for any optimal estimator: $\Delta\check{O}(d)^2 = 1/\mathcal{I}^Q_{\textsc{spe}}$. We also observe that the \textsc{sld} is a function of the source distribution and describes interference effects between different sources along the array. 
Surprisingly, the form of the estimator in Eq.~\eqref{eqn:optimal_estimator} has off-diagonal elements in the number basis, which suggests that intensity measurements along the near field is not the only optimal strategy. The existence of a second optimal estimator that is not photon number counting motivates an open question into the uniqueness of optimal measurements. A possible cause for this may be the degeneracy of the eigenstates of the generator.

%
%

\section{Conclusions and Discussions}
\label{sec:conclusions}

In this paper, we applied the theory of quantum estimation to an array of identical, stationary and equidistant emitters each with an intrinsic spatial Gaussian uncertainty profile. The quantum Fisher information (\textsc{qfi}) has been used as the figure of merit for the estimation of the source separation distance $d$ in the near field. We compare the estimation performance of different classical and quantum light sources. In order to efficiently report this comparison, we derive the generator $\mathcal{\hat{G}}$ responsible for changes in $d$ due to a general spatially homogeneous deformation $\xi$ applied to the array. These deformations change the expected mean positions of the sources leaving the source variances invariant. Each source was assumed to be mutually independent and was treated in their individual Hilbert spaces.

First, to quantify when the mutual independency of sources is valid and observe the dependence of the \textsc{qfi} on $d$, we allow for source overlaps. Calculating the \textsc{qfi} for an array of \textsc{spe}s, we found that a numerical approach is necessary to find the \textsc{qfi}. The \textsc{qfi} was observed to initially increases with $d$ until $d \sim \xi_ss/4$ after which it settles to the value consistent with those determined for $d \gg s$. In between these two regimes, we find that the \textsc{qfi} peaks slightly above that predicted when $d \gg s$. This is a nearest neighbour effect and remains for arbitrary $N$. It gave an indication that a source with higher average mode occupancy is favoured in this context for the estimation of $d$. With this insight, we explored the estimation performance of different classical and quantum states with the assumption of mutual independency.

We considered arrays of single photon emitters (\textsc{spe}s) as well as coherent, thermal and entangled light sources and conveniently summarise the following results
\begin{align}
\mathcal{I}^Q_{\textsc{e-spe}} = \frac{\xi_s^2 N(N^2-1)}{24s^2} = \frac{1}{2} \mathcal{I}^Q_{\textsc{spe}} = \frac{1}{4} \mathcal{I}^Q_{\text{Coh}} = \frac{1}{6} \mathcal{I}^Q_{\text{Bb}}.
\label{eqn:qfi_summary_results}
\end{align}
Interestingly, we find that the higher mode occupancies of the classical sources provide better estimates of $d$ than \textsc{spe}s. This is contrary to what may have been expected from earlier work where higher order correlations of \textsc{spe}s yield more information than thermal light sources. However, unlike this previous work, no post-selection of data was used here. The scalings determined here are based on the maximal information content in the state. The preference of classical sources in this context was found to be misleading. By using entanglement resource between the sources carefully, we find that it provides a precision enhancement. By constructing the optimal state which entangles the eigenstates corresponding to the maximum and minimum eigenvectors of the generator $\mathcal{\hat{G}}$, we found 
\begin{align}
\mathcal{I}^Q_\mathrm{Opt} = \frac{\xi_s^2 N^2 (N-1)^2}{4s^2}.
\label{eqn:optimal_state_qfi}
\end{align}
Physically the optimal state is found to be an extension of the \textsc{noon} state. It is a superposition of all $N$ photons emitted from the most extremal positions about the array centre $O$.

To address the optimal measurement scheme that saturates the \textsc{qcrb}, we first consider calculating the classical Fisher information for intensity measurements. Since we find it is equivalent to the \textsc{qfi}, photon number counting is found to be optimal. From the eigenbasis of the symmetric logarithmic derivative, we find the existence of a second optimal estimator. which is not photon counting To support our claim of its optimality, we confirm that it is unbiased and bounded by the quantum Fisher information. The existence of a multiple optimal estimators motivates an open question into the uniqueness of optimal measurements.

The considerations made in this report permit the precise evaluation of deformed coordinates of the quantum emitters and allows for corrective measures to negate their effects. This would find applications in evaluating stresses and strains and fracture prevention in materials expressing great sensitivities to deformations and to select a particular quantum source distinguished by its frequency from an array of reference or differing sources. Further research will consider the effect of incomplete detection of the state on the estimation precision, treatment of source efficiencies, temporal jitters, non-homogeneous deformations in higher dimensions and far field detection.

\section{Acknowledgments} 
\label{sec:acknowledgement}
This research was funded by the DSTL Quantum 2.0 Technologies Programme. The authors thank Mark Pearce for useful discussions.

\appendix
\section{The permutation group}
\label{sec:appendixA}

For some set $S_N$ of finite size $N$, the permutation group is the finite group $P$ whose elements are permutations of $S_N$. There exists $N!$ elements---each a bijection $\sigma: M \rightarrow M$. Hence, for $M = \{1, 2, \ldots, N\}$, each permutation can be written using Cauchy's two-line notation~\cite{Permutation_Groups_bk}
\[
\sigma = 
  \begin{pmatrix}
    1 & 2 & 3 & \cdots & N \\
    \sigma(1) & \sigma(2) & \sigma(3) & \cdots & \sigma(N)
  \end{pmatrix},
\]
which contains the elements of the set $S_N$ along the first row and the permutation image along the second. A cyclic notation is also commonly used. 
All finite groups can be represented as a group of permutations of a suitable set. In this work, when allowing for different sources to overlap, the contribution of different source pairings to the \textsc{qfi} becomes important. The permutation group is used to describe the combinatorics of this pairing. In the remainder of this section, we list identities associated with the mean source positions that were used when discussing the numerical results in \S~\ref{sec:diff_sources}. Let $\mu_j$ define the expected position of the $j^\text{th}$-source. Since each permutation is bijective,
\begin{align}
\sum_{j=1}^N  \mathfrak{G}\left[\mu_{\sigma(j)}\right] = \sum_{j=1}^N \mathfrak{G}\left[\mu_j\right],
\label{eqn:perm_result_1}
\end{align}
for some arbitrary well-defined operation $\mathfrak{G}$ on the elements of the permutation sets. A consequence of the same reasoning is the following
\begin{align}
\sum_{j=1}^N\mu_{\sigma(j)}'\mu_{\sigma(j)} = \sum_{j=1}^N\mu_j'\mu_j,
\label{eqn:perm_result_2}
\end{align}
where we recall $\mu_j' = \partial_d \mu_j$. Further, the evaluation of the \textsc{qfi} requires integrating over the continuous variable $x$ which runs along the source array. This operation becomes difficult to perform when the subscript contains a permutation element. To overcome this difficulty, we generate a `shift' property. For any group $G$ the inverse is a bijection of the set $S_N$. Hence, the inverse $\sigma^{-1} \in P$ such that for the sum over all permutations we have the following equality
\begin{align}
\sum_\sigma\sum_{j=1}^N\left[\mu_j'(x_{\sigma(j)} - \mu_j)\right] = \sum_\sigma\sum_{j=1}^N\left[\mu_j'(x_{\sigma^{-1}(j)} - \mu_j)\right].
\label{eqn:perm_result_3a}
\end{align}
From property~\eqref{eqn:perm_result_1} and the identity permutation $\sigma\sigma^{-1}(j) = \sigma^{-1}\sigma(j) = j$, we may re-write the \textsc{rhs} such that 
\begin{align}
\sum_\sigma\sum_{j=1}^N\left[\mu_j'(x_{\sigma(j)} - \mu_j)\right] = \sum_\sigma\sum_{j=1}^N\left[\mu_{\sigma(j)}'(x_j - \mu_{\sigma(j)})\right],
\label{eqn:perm_result_3b}
\end{align}
where we notice that the subscript $\sigma$ is shifted from the integration variable. Finally, for the form of the source positions used in this report we have
\begin{align}
\sum_\sigma\left[\sum_{j=1}^N \mu_j \mu_{\sigma(j)}\right] = 0.
\label{eqn:perm_result_4}
\end{align}
This may be seen by changing the order of the summation and that for each $j$ there are $(N-1)!$ terms with $\sigma(j) = k$ such that 
\begin{align}
(N-1)!\sum_{j=1}^N \mu_j \sum_{k=1}^N \mu_{k} = 0.
\label{eqn:perm_result_5}
\end{align}
The properties identified in Eqs~\eqref{eqn:perm_result_1}-\eqref{eqn:perm_result_5} were used in \S~\ref{sec:diff_sources} where we derived the dependence of the \textsc{qfi} with the source separation distance $d$. There, the permutation symbol naturally arose to describe all of the possible source pairings along the array.

\section{Normal ordering method}
\label{sec:appendixB}

For non commuting operators, there exists an ambiguity in the definitions of operator functions in quantum mechanics. The normal ordered form of a boson operator, where all creation operators appear to the left of annihilation operators, was developed to address the operator ordering ambiguity. There exist two well defined procedures on the boson operators which yield a normally ordered form; the normal ordering $\mathcal{N}$ operation and the double dot $: :$ operation.

For a general boson operator string $F(\hat{a}, \hat{a}^\dagger)$, normal ordering by means of the former method is achieved by repeated use of commutation relations until all creation operators appear to the left of annihilation operators. Under this operation the operator string remains the same but with a changed functional appearance. In the double dot operation, the normally ordered form of $F(\hat{a}, \hat{a}^\dagger)$ is achieved by assuming the creation and annihilation operators commute. The operator is in general changed with an equivalence maintained only if the operation is performed on a normally ordered string: $\mathcal{N} F(\hat{a}, \hat{a}^\dagger) = \; :\mathcal{N}F(\hat{a}, \hat{a}^\dagger):$

The normal ordering problem is solved when the following is satisfied~\cite{Blasiak2007_AJP}
\begin{align}
F(\hat{a}, \hat{a}^\dagger) = \mathcal{N}F(\hat{a}, \hat{a}^\dagger) = :F(\hat{a}, \hat{a}^\dagger):.
\label{eqn:normal_order_problem}
\end{align}
The two can be seen to have explicitly different meanings. A systematic approach to address all of the combinatorics associated with the latter form is Wicks theorem. A wide variety of numerical packages provide ease to this difficulty for polynomial expressions. However, this becomes increasingly cumbersome for increasing operator string lengths and the computational time grows exponentially. The normal ordering problem for non trivial operator strings of arbitrary lengths is an open area of research. 

We use the Bargmann representation to approach the normal ordering problem~\cite{Bargmann1961_CPAM}. It converts a boson operator string into one of multiplicative factors of a formal dummy variable, $\eta$, and its derivative by making the following transformations
\begin{align}
\hat{a}^\dagger_i \rightarrow \eta_i \quad \text{and} \quad \hat{a}_j \rightarrow \frac{\partial}{\partial\eta_j}
\label{eqn:bargmann_rep}
\end{align}
The commutation relation is preserved and this transformation makes the evaluation of vacuum expectation values of boson operator strings easier. Under this transformation map, the action of the annihilation operator $\hat{a}(k_j)\ket{0}=0$ is reproduced if the vacuum $\ket{0}$ maps to unity: $\partial/\partial\eta_j \cdot 1 = 0$. We use this representation to derive some results that have been used in the paper.

In section~\ref{subsec:spes} we determined the dependence of the \textsc{qfi} on the source separation distance by allowing relaxing the assumption that sources are mutually independent. The following vacuum expectation value was encountered.
\begin{align}
\langle0\vert\prod_{j=1}^N \hat{b}(x_j)\prod_{j=1}^N \hat{b}^\dagger(x'_j)\vert0\rangle = \sum_\sigma\prod_{j=1}^N \delta\left(x_j - x'_{\sigma(j)}\right).
\label{eqn:identity1}
\end{align}
The permutation sum identifies all the possible combinations of source overlaps along the array. This expression becomes increasingly cumbersome to determine for more complicated boson strings as encountered when using the generator to determine the \textsc{qfi}. Hence, to simplify the evaluation after this, we assumed mutual independence for different sources. By use of the Bargmann representation, we derive the following two expectation values:
\begin{widetext}
\begin{equation}
\langle0\vert \hat{b}_j(k'_j)^{n_j} \hat{n}_j(k''_j) \hat{b}_j^\dagger(k_j)^{n_j}\vert0\rangle = n_j! n_j \delta\left(k'_j - k''_j\right) \delta\left(k''_j - k_j\right) \delta\left(k_j - k'_j\right)^{n_j - 1},
\label{eqn:identity2}
\end{equation}
and
\begin{equation}
\langle0\vert \hat{b}_j(k'_j)^{n_j} \hat{n}_j(k''_j)\hat{n}_j(k'''_j) \hat{b}_j^\dagger(k_j)^{n_j}\vert0\rangle = n_j! n_j^2 \delta\left(k'_j - k''_j\right) \delta\left(k''_j - k_j\right) \delta\left(k_j - k'_j\right)^{n_j - 1},
\label{eqn:identity3}
\end{equation}
\end{widetext}
These vacuum expectation values are used to evaluate the variance of the generator. Specifically, the absence of the summation over all permutations reflects the use of updated commutation relations, which treat each source as distinct and mutually independent.



%

\end{document}